\documentclass[%
 reprint,
 amsmath,amssymb,
 aps,
]{revtex4-1}

\usepackage{graphicx}
\usepackage{dcolumn}
\usepackage{bm}
\usepackage{amssymb}
\usepackage{xspace}

\usepackage{color}

\begin{document}
\newcommand{\pp}{\ensuremath{\rm pp}\xspace}
\newcommand{\ppb}{\ensuremath{\rm p\!-\!Pb}\xspace}
\newcommand{\cme}{\ensuremath{\sqrt{s}}\xspace}
\newcommand{\pt}{\ensuremath{p_{\rm{T}}}\xspace}
\newcommand{\gev}{\ensuremath{{\rm GeV}/c}\xspace}
\newcommand{\pbpb}{\ensuremath{\rm Pb\!-\!Pb}\xspace}
\newcommand{\auau}{\ensuremath{\rm Au\!-\!Au}\xspace}
\newcommand{\ptopi}{\ensuremath{({\rm p+ \bar{p}}) / (\pi^{+}+\pi^{-})}\xspace}
\newcommand{\ktopi}{\ensuremath{({\rm K}^{+}+{\rm K}^{-}) / (\pi^{+}+\pi^{-})}\xspace}
\newcommand{\twotwo}{\ensuremath{2\rightarrow 2}\xspace}
\newcommand{\meanpt}{\ensuremath{\langle p_{\rm T} \rangle}\xspace}
\newcommand{\ltok}{\ensuremath{({\rm \Lambda}^{0}+\bar {\rm \Lambda}^{0})/(\rm 2 K^{0}_{s} )} \xspace}
\newcommand{\phitopi}{\ensuremath{(\rm 2 \phi)  / (\pi^{+}+\pi^{-})} \xspace}

\preprint{APS/123-QED}

\title{Color reconnection and flow-like patterns in \pp collisions}

\author{A. Ortiz Velasquez}%
 \email{antonio.ortiz$_$velasquez@hep.lu.se}
\author{P. Christiansen}
\affiliation{%
Lund University, Department of Physics, \\
Division of Particle Physics \\
Box 118, SE-221 00, Lund, Sweden.
}%

\author{E. Cuautle Flores}
 \altaffiliation[Also at ]{Benem\'erita Universidad Aut\'onoma de Puebla, Puebla, M\'exico}
\author{I. A. Maldonado Cervantes}%
\author{G. Pai\'c}%
\affiliation{%
 Instituto de Ciencias Nucleares, Universidad Nacional Aut\'onoma de M\'exico \\
 Apartado Postal 70-543, M\'exico Distrito Federal 04510, M\'exico. 
}%

\date{\today}

\begin{abstract}

Increasingly, with the data collected at the LHC we are confronted with
the possible existence of flow in \pp collisions. In this work we show
that PYTHIA 8 produces flow-like effects in events with multiple hard
subcollisions due to color string formations between final partons from
independent hard scatterings, the so called color reconnection. We present
studies of different identified hadron observables in \pp collisions at
7 TeV with the tune 4C. Studies have been done both for minimum bias and
multiplicity intervals in events with and without color reconnection to isolate the flow-like effect.

\end{abstract}                              

\maketitle

Since many years the existence of something that is generically called flow has been taken as a proof of collective behaviour of partons and hadrons.  According to this interpretation in central heavy ion collisions the transverse
pressure gradient causes a transverse hydrodynamic
expansion~\cite{bjorken:1}.  On the basis of hydrodynamic calculations the effect is transmitted to hadrons via a boost by the local velocity field, $u^\mu$~\cite{coop:1}. The transverse flow shifts the emitted particles to higher momenta,  and this effect increases for  heavier particles, because they gain more momentum from the flow velocity. It is important to stress that particles are not used for calculations, and a local thermal equilibrium is assumed as the starting point of the  hydrodynamic evolution of the system. Therefore, hydrodynamics does not tell us what is flowing and how these underlying degrees of freedom hadronize. Among the problems of the hydrodynamical models are the initial temperature, thermalization and the size of the system. Extending the hydrodynamical picture to \pp and \ppb~\cite{pp:flow:1,pp:flow:2,pp:flow:3}  brings us in contradiction with the basic tenet of hydrodynamics, {\it i.e.,} the mean free path of partons must be smaller than the size of the system.

\begin{figure} 
  \begin{center}
    \includegraphics[width=0.9\columnwidth]{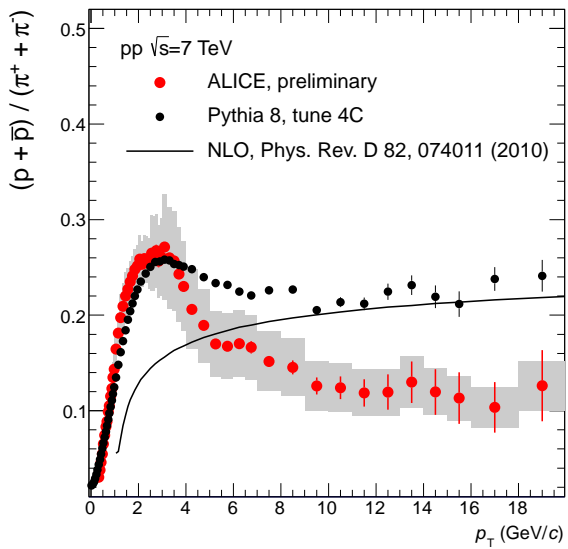}
    \caption{(Color online) Proton to pion ratio from \pp collisions at \cme $=$ 7
      TeV. ALICE data  are compared to results from PYTHIA 8 Tune 4C,
      as well as NLO QCD calculation~\cite{pia}.}
  \label{fig:pTopi:datatoMB}
\end{center}
\end{figure}

In heavy nuclei collisions the proton to pion  ratio exhibits an enhancement for transverse momentum (\pt) below 8 \gev and the position of the peak is pushed to higher momenta when one goes from peripheral to central \pbpb collisions~\cite{alice:pidPbPb}.  Surprisingly, Fig.~\ref{fig:pTopi:datatoMB} shows that in \pp collisions at $\sqrt{s}=$ 7 TeV we also observe a small enhancement  around 3 GeV/$c$~\cite{ortiz:qm12} which is qualitatively well reproduced by PYTHIA version 8.17~\cite{pythia8:0} tune 4C~\cite{pythia8:1}. For higher \pt ($> 8$ GeV/$c$) the description is poor, but for us this is not a major worry since we just use PYTHIA as a framework and we do not intend to tune it. In this letter our aim is to show that in PYTHIA the enhancement is attributed to color reconnection (CR). We argue that CR is another mechanism of flow where the boost is introduced at the partonic state just before hadronization in events with several multi-partonic interactions (MPI). The CR mechanism, which was originally introduced in PYTHIA~\cite{pythia6:2}, is microscopic and does not require a medium to be formed. This flow mechanism is very important because it could provide  an explanation of the observed flow-like patterns in \pp collisions~\cite{kisiel} and the collective phenomena seen in \ppb collisions~\cite{pPb:1}.

Note that here we shall only show evidence for radial flow, but the discussion will mention how higher order flow, {\it e.g.} elliptic flow, could be produced by the same mechanism.

 \begin{figure*}
\begin{center}
    \includegraphics[width=.88\textwidth]{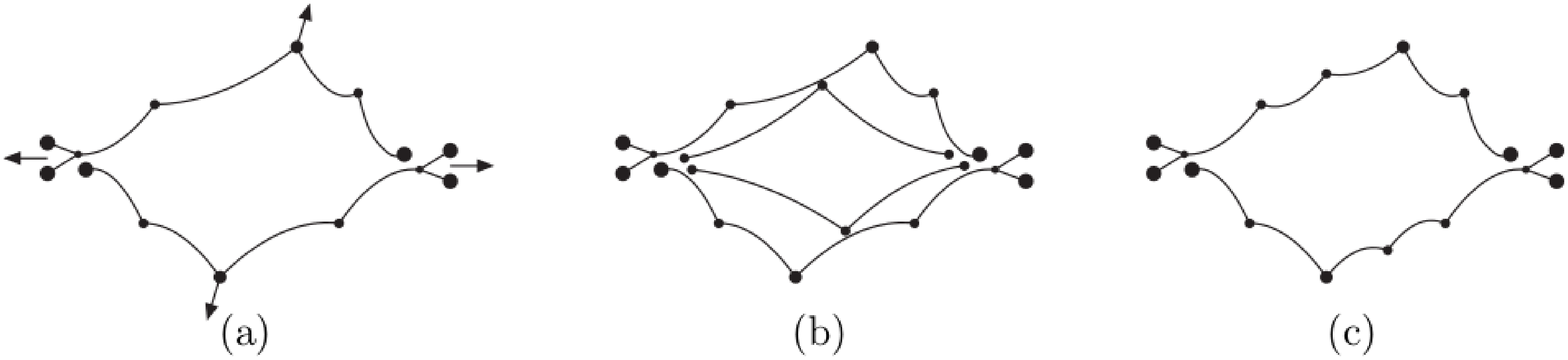}
    \caption{Illustration of the color reconnection in the string fragmentation model (picture taken from~\cite{cr:gosta}).  
      The outgoing gluons color connected to the projectile
      and target remnants (a). The second hard scattering  (b). Color reconnected string(c).}
  \label{string}
\end{center}
\end{figure*}
 

In hadron-hadron interactions it is possible to have multiple parton-parton
interactions in the same event, because beam-particles contain a
multitude of partons which can interact~\cite{cr:0}.  This is expected to happen in most
hadronic collisions at high energies, since, for example, in \pp collisions at LHC energies the inclusive jet cross section for $p_{\rm Tmin}$ below $\approx 4-5$ GeV exceeds the total cross section~\cite{pp:ue}. The inclusion of multi-parton interactions has been
supported by many experimental 
results~\cite{afsue, ua2ue, cdfue}. This is an important assumption used in
PYTHIA~\cite{mpi:1,pythia6:2} which allows to have a qualitatively good
description of the multiplicity distributions as well as the correlation of
observables like transverse sphericity with multiplicity in minimum bias (MB) \pp collisions
at the LHC energies~\cite{alice:st}.
But there are other issues, like the increase of the average transverse momentum with the multiplicity, which can be described through the so-called color reconnection.

Color reconnection was  first studied in a quantitative way,  in the context of rearrangements of partons at the perturbative level~\cite{cr:gosta:0}. More recently, different approaches to deal with CR have been developed~\cite{cr:2,cr:1, pythia8:1}. 
All these models are based on the calculation of the probability to connect partons by color lines. The models of CR include the probability to join low and  high \pt partons,  so the mechanism is present in soft and hard QCD processes.  However, at the perturbative level, CR is strongly suppressed~\cite{fadin}.

Fig.~\ref{string} shows a
sketch of CR in the string fragmentation model as implemented in PYTHIA where final partons are color connected in such a way that the total string length becomes as short as
possible~\cite{cr:gosta}. Therefore, the fragmentation of two independent hard scatterings are dependent and induces the rise of $\langle p_{\rm T} \rangle$ with
multiplicity.  However, in this work we discuss another feature of the model. In PYTHIA one string connecting two partons follows the movement of the partonic endpoints. The effect of this movement is a common boost of the string fragments (hadrons). Without CR, for a parton being  ``knocked out'' at mid-rapidity, the other string end will be part of the remaining proton moving forward so the boost is small (Fig.~\ref{string}a). With CR, 2 partons from independent hard scatterings at mid-rapidity can color reconnect and make a large transverse boost (Fig.~\ref{string}c). The last effect becomes more important in single events having several partonic subcollisions. This boost effect is similar to how flow affects hadrons in hydrodynamics, but the origin of the boost is clearly different in CR compared to hydrodynamics.

In order to pin down the effect, we calculated different baryon to meson
ratios as well as heavy meson to light meson ratios with and without CR. The
results were produced using primary charged particles defined as all final
particles including decay products except those from weak decays of strange
particles. This definition is similar to the one adopted by
ALICE~\cite{alice1}. The identified particle yields were computed at
mid-rapidity, $|y|<1$, and the event multiplicity in $|\eta|<2.4$. The
parameter which controls CR is the reconnection range, $\rm RR$, which
enters in the probability to merge a hard scale $\hat{p}_{\rm T}$
system with one of a harder scale, $(2.085\times {\rm
  RR})^{2}/((2.085\times {\rm RR})^{2}+\hat{p}_{\rm T}^{2})$. The tune
4C uses the value $\rm RR=1.5$ which gives a good description of
$\langle p_{\rm T} \rangle$ as a function of
multiplicity~\cite{pythia8:1}.

\begin{figure}[t!]
    \includegraphics[width=0.9 \columnwidth]{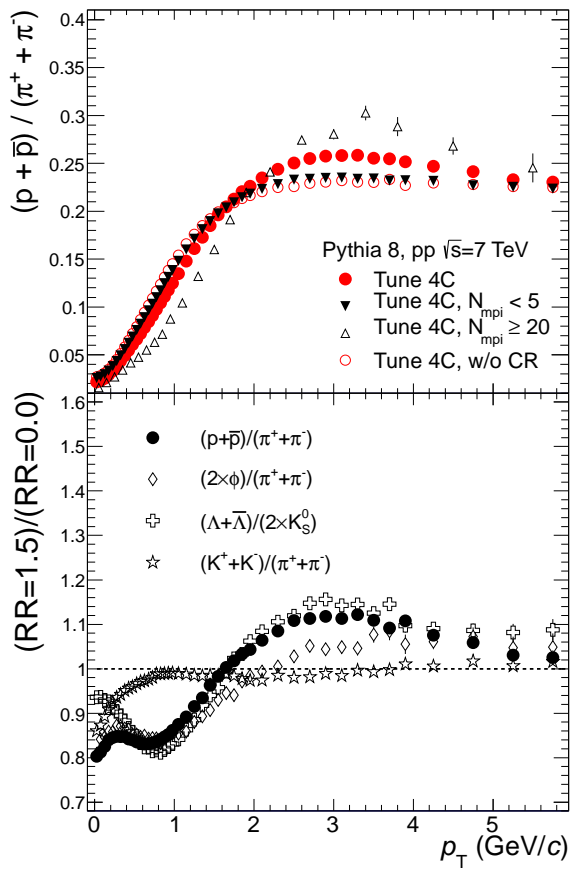}
    \caption{(Color online) Top panel: \ptopi as a function of \pt in
      \pp collisions simulated with PYTHIA 8 (solid circles), the ratio
      for events with low (solid triangles) and high (empty triangles)
      number of multi-parton interactions are overlaid. Results
      without color reconnection ($\rm RR=0.0$) are also shown (empty
      circles). Bottom panel: double particle ratios as a function of
      \pt for different hadron species.} 
  \label{fig:pTopi:ratios}
\end{figure}

Top panel of Fig.~\ref{fig:pTopi:ratios} shows the \ptopi ratio in the \pt-interval, $\pt <6$ \gev, for MB (PYTHIA 8 tune 4C) \pp collisions at 7
TeV. The distribution shows a clear bump around 2.5 GeV/c. The result
for simulations using $\rm RR=0$ (without CR) indicates a completely
different behaviour, for \pt larger than $\approx 1.8$ $\gev$ the
ratio stays flat. 
Therefore the origin of the peak is attributed to CR. The result also indicates that for \pt$>5$ GeV/$c$, the particles inside hard jets are not sensitive to color reconnection. The peak is enhanced if
one considers events with increased MPI activity. In the figure we plotted the
results for events with more than 20 MPI's, in this case the peak is also
pushed up to higher \pt, a characteristic effect of
  flow. The curve crosses the MB curve at $\approx$1.8 \gev
and the maximum ratio reaches 0.3 at 3 \gev. On the contrary, in events with
less than 5 MPI's the ratio behaves like in simulations without CR. The
evolution of the ratio with the number of MPI's is qualitatively similar to
the behaviour of the ratio from peripheral to central \pbpb
collisions~\cite{ortiz:qm12}. To study the effect of CR, the bottom panel of
Fig.~\ref{fig:pTopi:ratios} shows the double ratios with and without CR. For
\ptopi and \ltok one sees a bump at $\approx$ 2.5-3 \gev, similar to the one observed in the proton-to-pion ratio (top of the figure). The behaviour of the double ratios indicates that we have a mass effect since the $\phi/\pi $ and the \ptopi ratios exhibit a much larger bump than the \ktopi ratio. Even though in general PYTHIA underestimates the production of strange particles~\cite{alice1,strange:2}, we observe a 
hierarchy in the effect, it  increases with the  hadron mass, in a
pattern reminiscent of the radial flow in heavy ion collisions. To distinguish it from the flow usually associated with hydrodynamic evolution  we call it flow-like. We also observe the decrease in the double ratios at higher momenta, a feature that seems to be proper of CR and not of hydrodynamical behaviour.

\begin{figure}[t!]
    \includegraphics[width=0.9 \columnwidth]{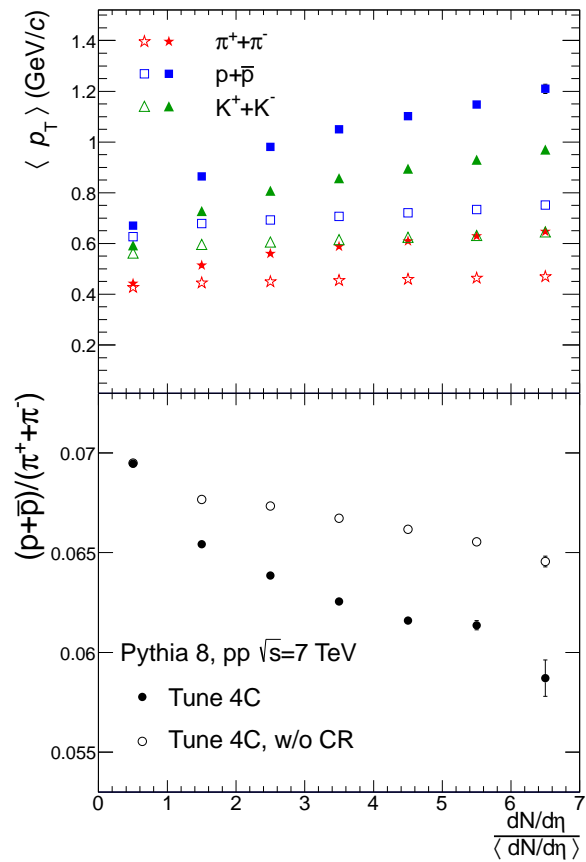}
    \caption{(Color online) Top panel: mean \pt as a function of the scaled event
      multiplicity for pions, kaons and protons. Results with CR (solid
      markers) and without CR (empty markers) are shown. Bottom panel:
      \pt-integrated proton-to-pion ratio as a function of the scaled
      event multiplicity. } 
  \label{fig:meanpt}
\end{figure}

At LHC energies the CMS Collaboration has published \pt spectra for
pions, kaons and protons as a function of the track
multiplicity~\cite{cms:1}. They found that \meanpt for protons
increases from $\approx$0.6 GeV/c to $\approx$1.4 \gev from their
lowest multiplicity class to the highest one. The upper
panel of Fig.~\ref{fig:meanpt} shows that this can only be accommodated
by PYTHIA simulations when color reconnection is included. In the
plot we observe an increase of \meanpt with multiplicity when the
color reconnection is turned on, while it looks flatter when color
reconnection is turned off. This is the expected behaviour, that is
used in PYTHIA to tune the amount of CR. However, our work offers an
interpretation from a different point of view; with CR the \meanpt of
protons increases faster than the pion one {\it i.e.} the effect
increases with the hadron mass. This observation is consistent with
the idea of the flow-like effect of string boosts.

We also made a combined fit of the pion, kaon and proton \pt spectra with a blast-wave function~\cite{bw:func}. From this fit one usually extracts the freeze-out temperature, $T_{\rm kin}$, and the average transverse velocity, $\langle \beta_{\rm T} \rangle$. We found
in CMS and PYTHIA data a  $T_{\rm kin}$-$\langle \beta_{\rm T} \rangle$ behaviour as a function of multiplicity very similar to the one observed in heavy ion collisions~\cite{alice:pidPbPb}.

Bottom panel of Fig.~\ref{fig:meanpt} shows that the \pt-integrated
\ptopi ratio decreases with the event multiplicity. It is interesting
that \pbpb data at the LHC exhibits a similar behaviour; a model
assuming a baryon-antibaryon annihilation does the qualitatively best
description of the trend~\cite{alice:pidPbPb}. In PYTHIA 8 this effect
is  caused by a change of the particle distribution in the phase
space. Actually without any cut on $y$ the $\bar{\rm p}/\pi$ ratio
stays constant as a function of multiplicity with or without CR. 

Finally, we want to say that the study was only focused on radial flow. However, we stress that the quantity used to minimize the string length $l=\sqrt{\Delta\eta^2 + \Delta\phi^2}$ has a dependence on the azimuthal opening angle, $\Delta\phi$, between the two partons. For radial flow we can understand that the partons from 2 independent interactions selects a preferred rapidity (minimizing $\Delta\eta$) and therefore boosts in that direction. In a similar way one could imagine that they could select a preferred $\phi$ giving rise to higher order flow, {\it e.g.}, elliptic and triangular. This point is very speculative and requires further theoretical investigations but here we want to point out that there is naturally a $\Delta\phi$ relation in the CR picture.


We have demonstrated that the CR scheme used in PYTHIA 8 event generator exhibits a  qualitatively new feature that has a potentially important consequence on our understanding of the details of the \pp collisions. 
The flow-like mechanism of PYTHIA does not require thermalization or a medium to be formed, and can hopefully lead to predictions that are different from hydrodynamics for small systems like \pp and ${\rm p-Pb}$.  We also note that the CR mechanism introduced in PYTHIA is microscopic and could potentially lead to novel hadronization effects if applied in a much denser environment such as heavy ion collisions. 

For the \ptopi ratio as a function of \pt (Fig.~\ref{fig:pTopi:datatoMB}) we have shown that even when hard
and soft processes are combined, such as in PYTHIA 8, one needs to introduce a
flow-like effect to obtain a peak. This suggest that this peak is a direct
indicator of a flow-like behaviour, {\it i.e.}, collective phenomena. We also remark that CR produces the mass dependent rise of $\langle \pt \rangle$ as a function of multiplicity therefore radial flow is an important effect which plays a role to explain the CMS results~\cite{cms:1}.

 
The authors acknowledge the very useful discussions with Torbj\"orn Sj\"ostrand on the details of PYTHIA 8. Support for this work has been received by CONACyT under the grant numbers 103735 and 101597; and PAPIIT-UNAM under the projects: IN105113 and IN107911. P.C. and A.O. acknowledge the support of the Swedish Research Council. The EPLANET fellowships have facilitated the necessary meetings in  this work in Mexico and at CERN.

\nocite{*}

\bibliography{apssamp}

\end{document}